# Comment on "Novel Public Key Encryption Technique Based on Multiple Chaotic Systems" [Phys. Rew. Lett. 95, 2005]

Wang Kai, Pei Wen-jiang, Zou Liu-hua, He Zhen-ya

(Department of Radio Engineering, Southeast University, Nanjing 210096, China)

Recently, a novel public key encryption technique based on multiple chaotic systems has been proposed [Phys. Rew. Lett., 95(9): 098702, 2005]. This scheme employs m-chaotic systems and a set of linear functions for key exchange over an insecure channel. The security of the proposed algorithm grows as $(NP)^m$, where $N$, $P$ are the size of the key and the computational complexity of the linear functions respectively. In this paper, the fundamental weakness of the cryptosystem is pointed out and a successful attack is described. Given the public keys and the initial vector, one can calculate the secret key based on Parsevala's theorem. Both theoretical and experimental results show that the attacker can access to the secret key without difficulty. The lack of security discourages the use of such algorithm for practical applications.

PACS numbers: 05.45.Vx, 89.70.+c

Chaotic systems are characterized by ergodicity, sensitive dependence on initial conditions and random-like behaviors, properties which seem pretty much the same required by several cryptographic primitive characters such as "diffusion" and "confusion". So far, the constructing of secret-key cryptosystems, which are mostly based on both chaotic synchronization and chaos-based pseudorandom bit generator, has attracted a great deal of attention, and plenty of chaos-based stream ciphers and block ciphers had been presented in the past [2, 3].

While all the currently used public-key cryptosystems based on number theory work well for both encryption and digital signatures, it is of much importance to construct public-key cryptosystems by chaotic dynamics and there have been some attempts along this line. In Ref. [4, 5], a public key cryptosystems and modification, which are based on distributed dynamics encryption, had been proposed. In those schemes, a high-dimensional dissipative dynamical system is separated into two parts. The binary message is characterize by different attractors, which are named as the 0-attractor and the 1-attractor, in the whole system. An authorized receiver knows the full dynamics and can simulate the system a priori in order to find the state space location of the two different attractors. This receiver is able to decode the message by observing the convergence of the system trajectory to the 0-attractor or to the 1-attractor in a reconstructed phase space. In Ref. [6], a RSA like public key algorithm based on the semigroup property of the Chebyshev map and the Jacobian Elliptic Chebyshev Rational map had been presented, and this algorithm had been used to design the key agreement scheme, deniable authentication scheme and Hash function [7]. Unfortunately, the later studies show that this scheme is vulnerable to some sophisticated attack because the semigroup property provides a lot of public keys from a public key with a fixed private key [8, 9], and the modification, designed along the same lines of the scheme, is not secure due to the same attack [10]. Furthermore, a Diffie-Hellman public-key cryptosystem named KKK protocol, which combines neural cryptography with chaotic synchronization, had be presented [11]. Compared with traditional supervised learning, there is no fixed target function in mutual learning scenario because each of the communicating parties acts as teacher and student simultaneously. Both of the two parities' statuses are chaotic which are driven by the random input. Although analytical and simulation results show that this scheme is vulnerable to genetic attack, geometric attack and probabilistic attack, this scheme is still a new and unexplored idea which makes it possible to use new types of cryptographic functions which are not based on number theory because the scheme can exchange a finite number of bits and generate very long keys by fast calculations [12]. Recently we introduced a scheme to improve the security by separating the hidden layer into two functionally independent units. Various security analyses demonstrate that the success probability for the most successful attack strategy, majority flipping attack, is $2^{-128}$. Meanwhile, we can obtain one bit of secret key after exchanging about three bits of information on average [13].

Recently, a novel public key encryption technique based on multiple chaotic systems, which can be seen as the mechanical analog of Diffie-Hellman protocol, has been proposed [1]. This scheme uses multiple chaotic systems and a set of linear functions for key exchange over an insecure channel. Various security analyses demonstrate that one has to solve the Diffie-Hellman problem in order to break the system, and the only viable attack is the brute force attack. The security of the proposed algorithm grows as $(NP)^m$, where $N$, $P$ and $m$ refer to the size of the key, the computational complexity of the linear functions and the number of linear functions respectively.

In this paper, we show that such a cryptosystem, although efficient and practical, unfortunately, is not secure. We describe an attack that permits to recover the corresponding secret key from the public key and the start value based on the Parseval's theorem. The lack of security discourages the use of such cryptosystem for practical applications.

The public key encryption uses the "m-chaotic systems based pseudorandom number generator" [14], and the process is shown in Fig. 1. Let's take $m = 2$ as an example. Assume that there are two different one-dimensional chaotic maps $F_1$, $F_2$, let $x_1(0), x_2(0)$ be initial value, and $\{x_1(k)\}$ $\{x_2(k)\}$ denote the two chaotic orbits. Define a pseudorandom bit sequence $k_i$, where

$$k_i = \begin{cases} 1, & x_1(i) > x_2(i) \\ no\ output, & x_1(i) = x_2(i) \\ 0, & x_1(i) < x_2(i) \end{cases} \quad (1)$$

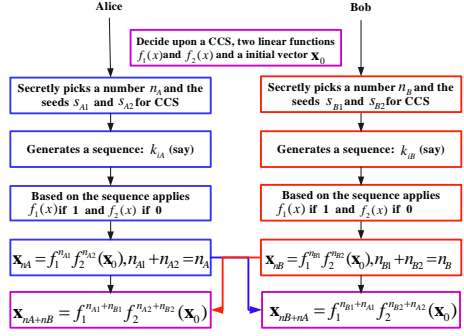

Fig.1. (color online). The proposed key exchange protocol using CCS.

The key exchange algorithm based on "a couple of chaotic systems based pseudo-random number generator" (CCS-PRNG) is described below.

Step 1: Alice and Bob, who wish to exchange the key, publicly agree on a common CCS, two linear functions $f_1(x)$ and $f_2(x)$, and a start vector $\mathbf{x}_0$.

Step 2: Alice secretly chooses two seeds $s_{A1}$ and $s_{A2}$ and a large number $n_A \in [0, N]$, and uses them as inputs to a CCS-PRNG to generate a random bit sequence $k_{iA}$ based on Eq. (1). $n_A$ is the number of bits in the sequence generated. This sequence $k_{iA}$ is provided as input to Eq. (2,3), along with $\mathbf{x}_0$ to iterate and give resultant $\mathbf{x}_{nA}$. The number $n_A$ can be varied to vary the security level of the system, as we shall show later. Alice publishes $\mathbf{x}_{nA}$ as her public key.

$$x_i = h(x_{i-1}, k_i) \quad (2)$$

$$h(x,k) = \begin{cases} f_1(x) & if \quad k = 0 \\ f_2(x) & if \quad k = 1 \end{cases} \quad (3)$$

Step 3: In a similar fashion, Bob secretly chooses two seeds $s_{B1}$ and $s_{B2}$ and a large number $n_B \in [0, N]$ to generate a bit sequence $k_{iB}$, to use in conjunction with $\mathbf{x}_0$ to generate his public key $\mathbf{x}_{nB}$.

Step 4: Alice uses Bob's public key $\mathbf{x}_{nB}$ as the seed and takes the same sequence, $k_{iA}$, as generated earlier, to perform another $n_A$ iteration to get secret key $\mathbf{x}_{nB+nA}$.

Step 5: Similarly, Bob uses Alice's public key as the seed and takes the same bit sequence, $k_{iB}$, as generated by him earlier, to perform another $n_B$ iteration to get secret key $\mathbf{x}_{nA+nB}$.

It should be noted that $f_1(x)$ and $f_2(x)$ are linear functions, i.e., they satisfy the condition $f_1 \circ f_2 = f_2 \circ f_1$, therefore the sequence of operation does not matter. Therefore, the resultant iterations of the two sets are equal, i.e., $\mathbf{x}_{nB+nA} = \mathbf{x}_{nA+nB}$. This value obtained can now be suitably mapped to a common key $K$, which can then be used to communicate over the insecure channel.

The following example illustrates the algorithm. Let the two linear functions be $f_1(x) = FFT(x)$, $f_2(x) = 1.5x$, the initial vector $\mathbf{x}_0 = [0.06, 0.35, 0.81, 0.01, 0.14]$. This is randomly chosen by one of the parties (Alice or Bob) and is made public. Let the CCS, used by both Alice and Bob, be $F_1(x_{n+1}) = 4x_n(1-x_n)$ and $F_2(x_{n+1}) = 3.98x_n(1-x_n)$. Alice secretly picks a number $n_A = 10$ and uses the seeds $\{0.83, 0.34\}$. Bob secretly picks a number $n_B = 12$ and uses the seeds $\{0.47, 0.61\}$. The random bit streams generated by the CCS are: $k_A = [1,0,1,0,1,10,1,0,1]$ and $k_B = [0,1,0,0,0,1,1,0,0,1,1,1]$. The resulting outputs published as the public keys are: $\mathbf{x}_{nA} = [3.80, 88.59, 6.33, 512.58, 221.48]$ and $\mathbf{x}_{nB} = [8.5, 199.3, 14.2, 1153.3, 498.3]$. The secret key $\mathbf{x}_{nA+nB} = [5410, 315360, 729820, 9010, 126140]$.

In order to break the system, one has to solve the Diffie-Hellman problem. If the linear functions $f(x)$ are suitably chosen, one cannot easily guess the constituent operations that convert the start vector $\mathbf{x}_0$ to the public keys $\mathbf{x}_{nA}$ and $\mathbf{x}_{nB}$, and then can not find $\mathbf{x}_{nA+nB}$ from $\mathbf{x}_{nA}$ and $\mathbf{x}_{nB}$. The only viable attack is the brute force attack, where the adversary has to try all the possible combinations of sequences. Let the value $\mathbf{x}_{nA}$ and $\mathbf{x}_{nB}$ be chosen in the range $[0, N]$. Let each of the linear functions $f_1(x), f_2(x), \cdots, f_m(x)$ require on the order of $P$ floating point operations to execute. Then, in order to establish the key (by Alice or Bob), it requires on the order of $NP$ floating point operations. However, the adversary has to decide for every number in the sequence which of the linear functions to use. Therefore, the complexity to break the cryptosystem is on the order of $(NP)^m$ [1].

The public-key cryptosystem was presumed to be secure based on the following observation: the adversary can not easily guess the constituent operations that convert the start value $\mathbf{x}_0$ to the public key $\mathbf{x}_{nA}$ and $\mathbf{x}_{nB}$. However, as shown in steps 2 and 4 of the algorithm, if the adversary calculates the total numbers of "1" and "0" in the bit sequence $k_{iA}$, he can calculate the secret key $\mathbf{x}_{nA+nB}$ based on the equation: $\mathbf{x}_{nA+nB} = f_1^{n_{A1}} f_2^{n_{A2}}(\mathbf{x}_{nB})$, where $n_{A1}$ and $n_{A2}$ are the total numbers of "1" and "0" in $k_{iA}$. In the following, we will take the linear functions $f_1(x) = FFT(x)$, $f_2(x) = 1.5x$ as example and show that the above cryptosystem is not secure. Given the public keys and the start value, the adversary can recover $n_{A1}$ and $n_{A2}$ after some algebra, and then the public key can be calculated correctly.

Let the initial vector $\mathbf{x}_0$ be a $m$-dimensional vector and $\mathbf{X}(k) = f_1(\mathbf{x}_0)$. The equation $M \sum_{n=0}^{M-1} |\mathbf{x}_0(n)|^2 = \sum_{k=0}^{M-1} |\mathbf{X}(k)|^2$ will be hold according to Parseval's theorem [15]. The equation $M^{n_{A1}} \sum_{n=0}^{M-1} |\mathbf{x}_0(n)|^2 = \sum_{k=0}^{M-1} |\mathbf{X}(k)|^2$ will be hold if $\mathbf{X}(k) = f_1^{n_{A1}}(\mathbf{x}_0)$, and $1.5^{2n_{A2}} \sum_{n=0}^{M-1} |\mathbf{x}_0(n)|^2 = \sum_{k=0}^{M-1} |\mathbf{X}(k)|^2$ if $\mathbf{X}(k) = f_2^{n_{A2}}(\mathbf{x}_0)$. When $\mathbf{X} = f_1^{n_{A1}} f_2^{n_{A2}}(\mathbf{x}_0)$, the equation $M^{n_{A1}}(1.5)^{2n_{A2}} \sum_{n=0}^{M-1} |\mathbf{x}_0(n)|^2 = \sum_{k=0}^{M-1} |\mathbf{X}(k)|^2$ will be hold for sure. Based on this fact, the attack strategy is described as follows:

Step 1: According to the start value $\mathbf{x}_0$, calculate $\sum_{n=0}^{4} |\mathbf{x}_0(n)|^2 = 0.7983$;

Step 2: According to the public key $\mathbf{x}_{nA}$, calculate $\sum_{n=0}^{4} |\mathbf{x}_{nA}(n)|^2 = 319695$, and then calculate

$M^{n_{A1}}(1.5)^{2n_{A2}} = 4.0045 \times 10^5$. Let $M^{n_{A1}}(1.5)^{2n_{A2}} = S$;

Step 3: Let's define two variables $I_{n1}$ and $I_{n2}$ firstly. We enumerate every possible integer $I_{n2}$ in $[0, N]$ to find the least value $I_{n2} = 4$, which makes $I_{n1} = \log_M C = 6$ a positive integer, where $C = S/(1.5)^{2I_{n2}} = 15625$. Let $n_{A1} = I_{n1}$, $n_{A2} = I_{n2}$;

Step 4: Calculate the secret key $\mathbf{x}_{nA+nB} = [5410, 315360, 729820, 9010, 126140]$ with the public key $\mathbf{x}_{nB}$ according to $f_1^{n_{A1}} f_2^{n_{A2}}(\mathbf{x}_{nB})$.

In conclusion, several weaknesses in such public key encryption technique are shown. Given the public keys and the initial vector, the adversary can recover the secret key correctly. A dynamic system with deterministic randomness may be used to improve the security of ergodicity based chaotic cryptosystem [16]. In fact a realizable asymptotic model to describe the deterministic randomness has been reported [17].

**Acknowledgements**

This work was supported by the Natural Science Foundation of China under Grant 60133010 and 60102011, the National High Technology Project of China under Grant 2002AA143010 and 2003AA143040, the Excellent Young Teachers Program of Southeast University.

**References**

[1]   B. Ranjan, Phys. Rew. Lett. 95, 098702 (2005).
[2]   M. S. Baptista, Phy. Lett. A 240, 50 (1998).
[3]   G. Jakimoski, L. Kocarev, IEEE Trans. Circuits I, 48(2), 163 (2001).
[4]   R. Tenny, L. S. Tsimring, L. Larson L and H.D.L. Abarbanel, Phy. Rev. Lett. 90, 047903 (2003).
[5]   R. Tenny, L. S. Tsimring, IEEE Trans. Circuits I, 52 (3), 672 (2005).
[6]   L. Kocarev, M. Sterjev, A. Fekete, et al., Chaos, 14 (4), 1078 (2004).
[7]   D. Xiao, X. F. Liao, K. Wong, Chaos, Soliton. Fract., 23(4), 1327 (2005).
[8]   T. Kohda, T. Yosimura, Proc. of ISCAS'04, 4: 573 (2004).
[9]   T. Kohda, T. Yosimura, Proc. of ISCAS'04, 2004, 4: 573-576
[10]  G. Alvarez, Chaos, Soliton. Fract., 26 (1), 7 (2005).
[11]  Kanter I, Kinzel W, Kanter E. 2002, 57(1):141-147; Mislovaty R, Klein E, Kanter I, et al. Phys. Rew. Lett., 2003, 91(11):118701; Ruttor A, Kinzel W, Shacham L, et al., Phys. Rew. E, 2004, 69(4):046110; Klein E, Mislovaty R, Kanter I, et al., Phys. Rew. E, 2005, 72 (1): 016214
[12]  A. Kilmov, A. Mityagin and A. Shamir, Lecture Notes In Computer Science, 2002, p. 288.
[13]  J. T. Zhou, Q. Z. Xu, W. J. Pei, et al, Int. J. Neural Systs. 14, 393 (2004).
[14]  S. J. Li, X. Q. Mou and Y. L. Cai, in Proc. INDOCRYPT'01, 2001, p.316.
[15]  A. V. Oppenheim and R. W. Schafe, Digital signal processing. Prentice—Hall Inc, 1975.
[16]  J. A. Gonzalez, L. I. Reyes and J. J. Suarez, Physica A 316, 259 (2002).
[17]  Q. Z. Xu, S. B. Dai, W. J. Pei, et al., Neur. Inform. Proc-Lett. Rev. 3, 21 (2004).